\documentclass{emulateapj} \usepackage{apjfonts} \usepackage[sort&compress]{natbib}

\shorttitle{Warm gas accretion}
\shortauthors{Murante et al.}

\begin{document}


\title{ A warm mode of gas accretion on forming galaxies}


\author{Giuseppe Murante}
\affil{Osservatorio di Torino,
    Strada Osservatorio 20, I-10025, Pino Torinese (TO), Italy}
\affil{Dipartimento di Fisica - Sezione di Astronomia, Universit\`a di Trieste, via Tiepolo 11, I-34131 Trieste -- Italy }
\email{murante@oato.inaf.it}

\author{Matteo Calabrese}
\affil{Dipartimento di Fisica Generale ``Amedeo Avogadro'', Universit\`a degli
  Studi di Torino, Via P. Giuria 1, I-10125, Torino (Italy)}
\email{calabrese@oato.inaf.it}

\author{Gabriella De Lucia}
\affil{I.N.A.F, Osservatorio di Trieste,
    Via Tiepolo 11, I- 34131, Trieste, Italy}
\email{delucia@oats.inaf.it}

\author{Pierluigi Monaco}
\affil{Dipartimento di Fisica - Sezione di Astronomia, Universit\`a di Trieste, via Tiepolo 11, I-34131 Trieste -- Italy }
\affil{I.N.A.F, Osservatorio di Trieste,
    Via Tiepolo 11, I- 34131, Trieste, Italy}
\email{monaco@oats.inaf.it}

\author{Stefano Borgani}
\affil{Dipartimento di Fisica - Sezione di Astronomia, Universit\`a di Trieste, via Tiepolo 11, I-34131 Trieste -- Italy }
\affil{I.N.A.F, Osservatorio Astronomico di Trieste,
    Via Tiepolo 11, I- 34131, Trieste, Italy}
\affil{I.N.F.N,  Sezione di Trieste, Trieste, Italy}
\email{borgani@oats.inaf.it}

\and

\author{Klaus Dolag}
\affil{University Observatory M\"unchen, Scheinerstr. 1, 81679,
  M\"unchen, Germany}  
\email{kdolag@mpa-garching.mpg.de}




\begin{abstract}
  We present results from high--resolution cosmological hydrodynamical
  simulations of a Milky--Way-sized halo, aimed at studying the effect
  of feedback on the nature of gas accretion. Simulations include a
  model of inter-stellar medium and star formation, in which SN
  explosions provide effective thermal feedback. We distinguish
  between gas accretion onto the halo, which occurs when gas particles
  cross the halo virial radius, and gas accretion onto the central
  galaxy, which takes place when gas particles cross the inner
  one-tenth of the virial radius. Gas particles can be accreted
  through three different channels, depending on the maximum
  temperature value, $T_{\rm max}$, reached during the particles' past
  evolution: a cold channel for $T_{\rm max}<2.5 \times 10^5$ K, a hot
  one for $T>10^6$K, and a warm one for intermediate values of $T_{\rm
    max}$. We find that the warm channel is at least as important as
  the cold one for gas accretion onto the central galaxy. This result
  is at variance with previous findings that the cold mode dominates
  gas accretion at high redshift. We ascribe this difference to the
  different supernova feedback scheme implemented in our
  simulations. While results presented so far in the literature are
  based on uneffective SN thermal feedback schemes
  and/or the presence of a kinetic feedback, our simulations include
  only effective thermal feedback.  We argue that
  observational detections of a warm accretion mode in the
  high--redshift circum-galactic medium would provide useful
  constraints on the nature of the feedback that regulates star
  formation in galaxies.
\end{abstract}





\section{Introduction}

In the current standard picture for structure formation, galaxies are
believed to form from condensation of gas within the potential wells
of dark matter halos (White \& Rees 1978). Therefore, understanding
how gas is accreted onto dark matter haloes and galaxies is crucial to
understand the process of galaxy formation.

Substantial numerical work recently focused on the `cold accretion' mode
(Birnboim \& Dekel 2003; Keres et al. 2005) as the main driver for gas
accretion at high redshift and for haloes less massive than a few times
$10^{11}\,{\rm M}_{\odot}$. The distinction between two modes of accretion,
cold and hot, was clearly understood when the first hierarchical galaxy
formation models were presented (Rees \& Ostriker 1977; Binney 1977; see also
Benson \& Bower 2011). If the halo virial temperature is larger than the
temperature of the accreting gas, this will accrete supersonically, which
causes the formation of an accretion shock. If the gas cooling time is longer
than the 
dynamical time of the halo, the shock will occur at a radius that is comparable
or slightly larger than the virial radius (Bertschinger 1985; Evrard 1990). If
the cooling time is much shorter than the halo dynamical time, the 
shock will occur much closer to the forming galaxy.  The gas is still heated to
very high temperature, but it cools so rapidly that it cannot maintain the
pressure needed to support a quasi-static hot atmosphere\footnote{In recent
  work, gas in cold flows is also collimated, i.e. it is accreted along cosmic
  filaments}.

This picture has been validated by 1D hydrodynamical
simulations (Birnboim \& Dekel 2003; see also Forcada-Miro \& White
1997), and by recent 3D hydrodynamical simulations (Keres et al. 2005;
Ocvirk et al. 2008). These simulations have shown that a static
atmosphere of hot, virialized gas develops in haloes above $2.3\times
10^{11} {\rm M}_{\odot}$, and that this `transition mass' is nearly constant
as a function of redshift. Simulations have also pointed out that,
even when a shock is present, cold gas accretion can occur along
filaments that penetrate deep inside the hot halo and might funnel gas
to the central galaxy, elevating its star formation rates.

Brooks et al. (2009) investigated gas accretion in a set of high resolution
simulations spanning about two orders of magnitude in halo mass. They find
that, for galaxies in the range $0.01\,L_\star$ to $L_\star$, the early growth
of the stellar disk is dominated by gas that is not shocked to the virial
temperature of the dark matter haloes, and that cold accretion dominates at
all times for galaxies less massive than our Milky Way. A high fraction of gas
accretion onto halos is found to be smooth, even for the most massive galaxy
considered. In a more recent study, van de Voort et al. (2011) used the
OverWhelmingly Large Simulations (OWLS; Schaye et al. 2010) to investigate
accretion rates onto haloes and galaxies, and their dependence on halo mass
and redshift. They confirm that hot mode accretion dominates the growth of
massive haloes, but show that accretion onto galaxies is very sensitive to
feedback processes and metal line cooling. In addition, they confirm that gas
accretion on the halo is smooth. 
Finally, Oppenheimer et al (2010)
studied how gas accretion is affected by kinetic feedback. They found that a
significant amount of the gas that later form stars has been previously
processed through the winds. This {\it wind channel} is relevant below $z
\approx 2$.

In this paper, we re-address the issue of gas accretion onto the halo and onto
the galaxy, taking advantage of high-resolution hydrodynamical simulations of
galaxy-size haloes. We use our new star--formation and feedback algorithm,
  MUPPI (Murante et al 2010), that includes an effective 
{\it thermal} feedback
  but no kinetic feedback.
Our aim is to verify whether, in this scheme, gas accretion onto
the central part of the halo remains mainly cold and smooth, and whether these
results are robust against increasing numerical resolution, well above that
adopted in cosmological simulations.

\section{Numerical Methods}
\label{sec:sim}

\subsection{Simulations}

We study the accretion of gas onto the {\it halo} and on its central part, the
{\it galaxy}, using hydrodynamical simulations of galaxy-size haloes. Initial
conditions were generated by Stoehr et al. (2002) adopting a $\Lambda$CDM
cosmology, with $\Omega_m=0.3$, $\Omega_\Lambda=0.7$, $H_0=70$ km s$^{-1}$
Mpc$^{-1}$ and $\sigma_8=0.9$.  A `Milky Way' halo was selected from an
intermediate-resolution simulation as a relatively isolated halo which
suffered its last major merger at $z > 2$, and with approximately the correct
peak rotation velocity at $z=0$. Our simulations include as many gas particles
as high-resolution DM particles, and adopt a baryon fraction of $f_b=0.19$. In
Table~\ref{table:ic}, we list the main numerical parameters of the
simulations. The highest resolution simulation (R2) was run down to $z=2.5$.

Our simulations were carried out using the Tree+SPH code GADGET-3, an improved
version of the GADGET-2 code (Springel 2005), by using the star formation and
feedback scheme MUPPI \citep[M10]{M10}. In this scheme, each star-forming gas
particle is described as a multi--phase particle, where cold and hot gas
co-exist in pressure equilibrium. Mass and energy fluxes between different
phases are described by a system of ordinary differential equations.  The
molecular fraction of cold gas is estimated using a phenomenological
prescription \citep{BlitzRosolowski06}. Stars explode and inject a fraction
$f=0.3$ of the available SNe energy as thermal energy in the hot and tenuous
gas phase of neighbours. These do not immediately radiate it away because of
the hot gas low density.  As a result, our scheme provides effective  
thermal
feedback: gas particles inside and near star--forming regions are heated, and
escape from their parent structure along the path of minimal resistance
(i.e. the direction of lowest density), with velocities up to $\approx 100$ km
s$^{-1}$, giving rise to galactic fountains.  We stress that we do not include
kinetic feedback.  A particle remains in the multi--phase stage for twice the
gravitational dynamical time it had when 95\% of its cold phase was
accumulated. This simulates the formation and disruption of star-forming GMC.
We refer to M10 for a detailed description of our scheme and for a discussion
of the choice of the relevant parameters, which we kept fixed here.  Our model
has been included in a recent code-comparison project on galaxy formation
simulations (Scannapieco et al, 2011).

Our simulations also include a phenomenological description of the UV
background (Haardt \& Madau 1996), but no chemical evolution of the gas: gas
cooling is computed assuming primordial chemical composition.

\begin{table}
  \caption{Simulations. Column 1: Simulation name. Column 2: Mass of the dark
    matter particle, in h$^{-1}$ M$_\odot$. Column 3: Mass of
    the gas particle, in h$^{-1}$ M$_\odot$. Column 4:
    Plummer-Equivalent softening length at $z=0$, in h$^{-1}$ kpc. The
    softening is physical down to redshift $z=2$, and comoving
    afterwards. Column 5: Number of DM and gas particles in the high-resolution
    region. Column 6: Virial mass of the halo, in h$^{-1}$
    M$_\odot$. Here we define the virial mass as that contained in a radius
    enclosing a mean overdensity of 200 times the critical density. The
    simulation R2 was run down to $z=2.5$ only.} 
\begin{tabular}{c c c c c c}
\hline\hline Name & $M_{\rm DM}$ & $M_{\rm Gas}$& $\epsilon$ & $N_{\rm
  DM}$ \& $N_{Gas}$&  $M_{\rm vir}$ \\
\hline
R1    & $1.6 \times  10^6$   & $3.0 \times  10^5$  & 0.325 & 5,953,033 & $2.70   \cdot 10^{12}$\\
R2    & $1.7 \times  10^5$  & $3.2 \times  10^4$ & 0.155 & 55,564,205 & - \\
\hline
\end{tabular}
\label{table:ic}
\end{table}

\subsection{Analysis method}

We analyze the thermal history of gas particle accreting {\it onto the
  halo}, and {\it onto the galaxy} by using the following definitions.
The {\it halo} is defined as a sphere centered on the most bound
particle, and with virial radius enclosing an overdensity of 200 times
the critical density at the redshift of interest. The {\it galaxy}
includes all particles within one tenth of the virial radius. We will
see that this distinction is very important, since the gas accreting
onto these components can have different properties. To perform our 
analysis, we used 131 snapshots, roughly equispaced in redshift from $z=8.3$. 

At each snapshot, we selected the most massive
progenitor of our $z=0$ halo, by using a standard friends-of-friends
algorithm with linking length $b=0.16$.  We then assumed that a gas
particle is accreted onto the halo (galaxy) at a snapshot
$n$ if it is part of the halo (galaxy) at the corresponding
snapshot but not at the previous one $n-1$.  Each gas particle is
then traced back in time, and the highest temperature reached during its
past evolution is recorded.

We distinguish gas accreting in three
channels: {\it(i)} cold gas, made of particles that never
reached temperatures larger than $T_{\rm min}=2.5 \times 10^5$ K; {\it
  (ii) warm} gas, composed of particles that were hotter than $T_{\rm
  min}$ but colder than $T_{\rm max}=10^6$ K; {\it (iii)} hot gas,
made of particles that reached temperatures larger than $T_{\rm
  max}$. In previous work, the {\it warm} channel has been included in the hot
one. In this study, we have considered this separate channel to quantify the
accretion that occurs at `intermediate' temperatures, that are indicative of
feedback from supernovae (SNe) rather than shock heating.

Finally, we identified all self-bound structures at each snapshot
using the algorithm {\small SUBFIND} (Springel et al. 2001), and built merger
trees for all of them using the same procedure described 
in Springel et al. (2005).

Using these merger trees, we check if gas particles
accreted on the halo or on the galaxy belong to a distinct
subhalo at the previous time, i.e. for each accretion channel, we determine the
fraction of gas accreted in 
``clumps''. Gas particle that were not assigned to a subhalo in the
past are considered as ``smoothly'' accreted.

\section{Results}
\label{sec:res}

\begin{figure*}
\hspace{2.5truecm}
\vbox{\hbox{\includegraphics[scale=0.60,angle=-90]{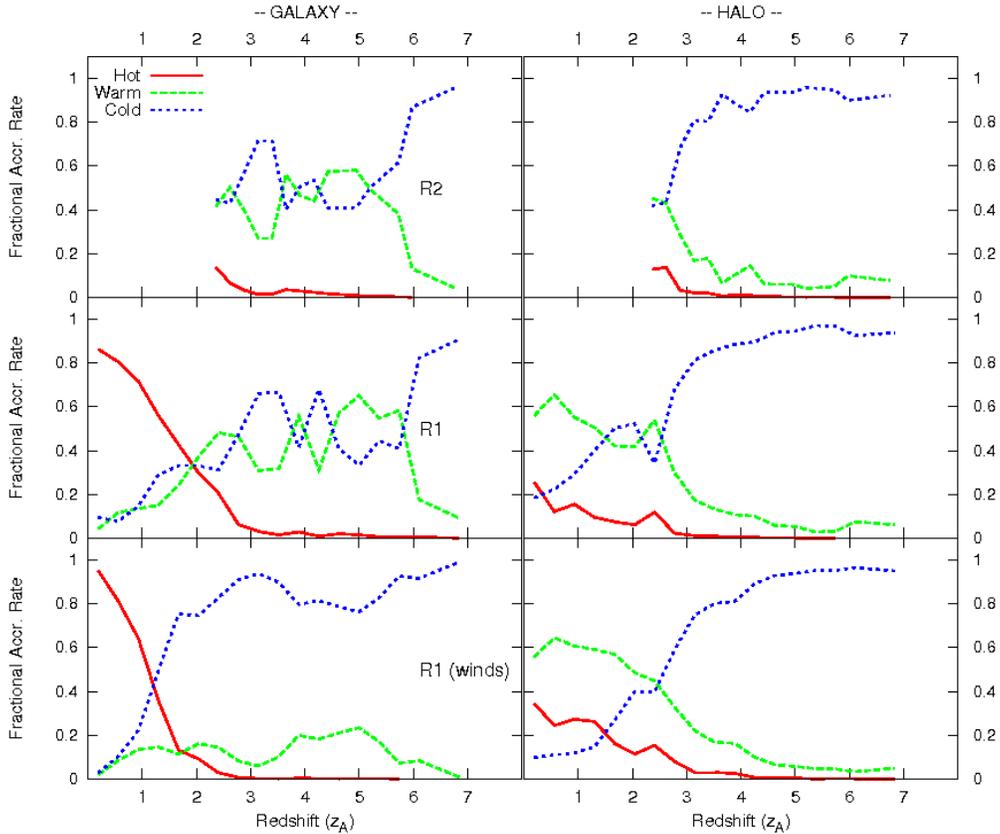}}}
\caption{Right panels: gas fractional accretion rates onto 
  the {\it halo}, as a function
  of the accretion redshift. Left panels: corresponding fractional gas 
  accretion rates
  onto the {\it galaxy}. In red, we show the hot gas accretion, in green the
  warm accretion, and in blue the cold accretion. In pink, we show the total
  gas accretion rate. For clarity, we have resampled our redshift outputs in 20
  redshift bins.
\label{fig:accr_channels}}
\end{figure*}

Figure~\ref{fig:accr_channels} shows the fractional mass accretion
rates in the three channels defined above. Right panels are for
accretion on the {\it halo}, while left panels show the corresponding
accretion rates on the {\it galaxy}. We show the accretion rates for
the resolution levels R1 and R2, as well as for a R1 run that adopted
the Springel \& Hernquist (2003) effective model
(see below).
The Figure shows that gas is
accreted on the halo mainly via the {\it cold} channel for all runs
considered.  At low redshift, $z<2$, the importance of {\it warm} and
{\it hot} channels increases, in agreement with previous studies
(as in, e.g., Keres et al. 2006, ,Brooks et al 2009, van de Voort et al 2011). 

Results are quite different for the accretion onto the galaxy. In this case,
the {\it cold} and the {\it warm} channels equally contribute to gas accretion
at high redshift.  At later times, the {\it hot} channel provides the dominant
contribution.

Our findings are robust against resolution: our R2 simulation has a mass
resolution 
which is sufficient to resolve
halos of $\approx 10^6$ M$_\odot$ with $\approx 100$ DM particles. Therefore,
results from this simulations are reliable also at high redshift and, as shown
in Figure~\ref{fig:accr_channels}, they are in excellent agreement with those
obtained for the R1 run.

\begin{figure*}
\hspace{2.5truecm} \vbox{ \hbox{
    \includegraphics[scale=0.5]{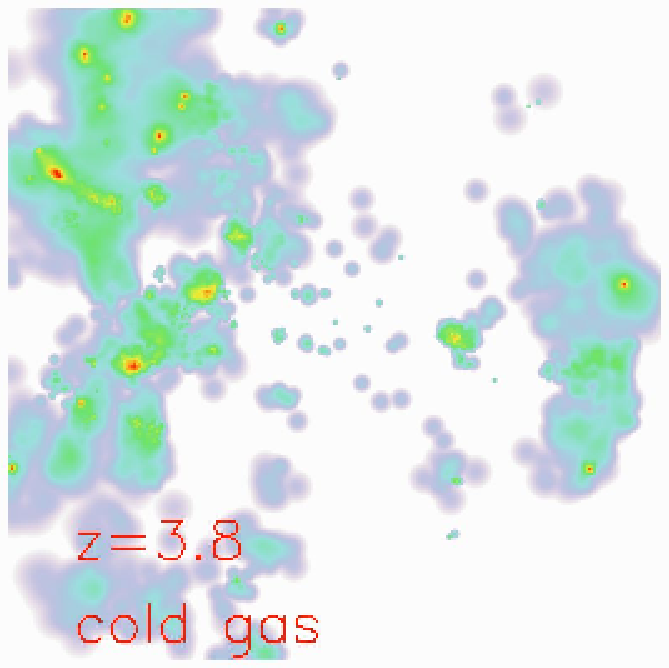}
    \includegraphics[scale=0.5]{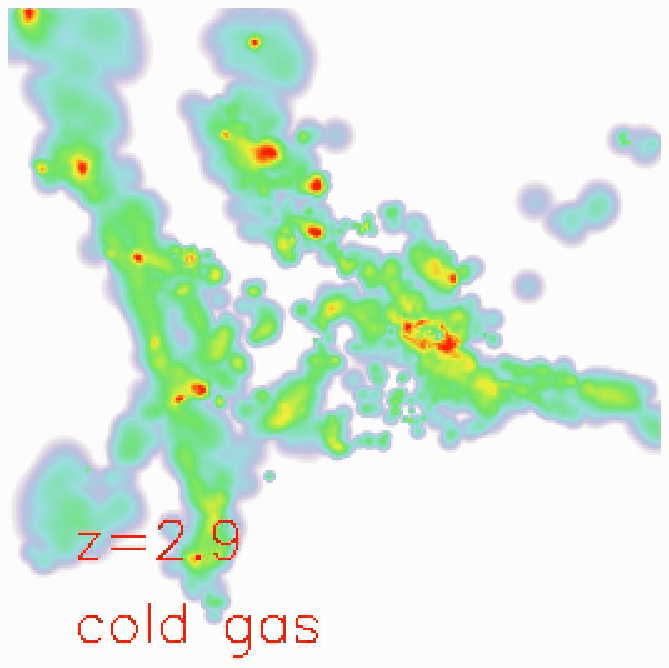}
    \includegraphics[scale=0.5]{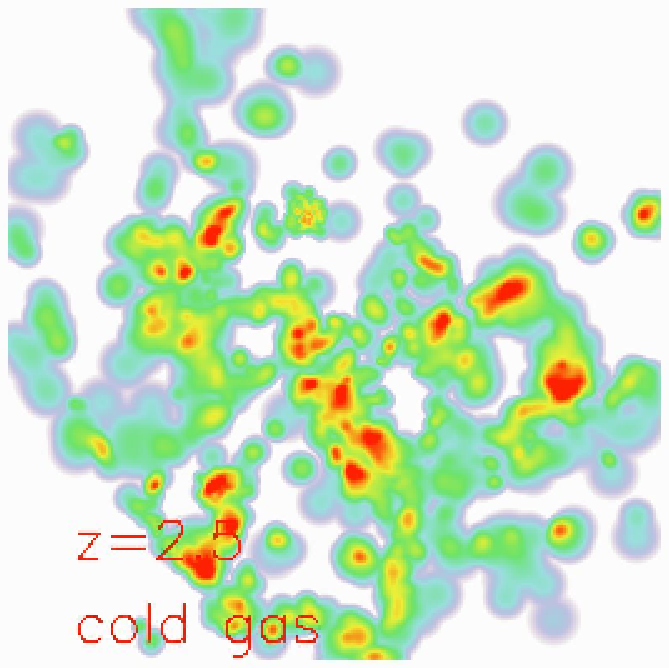} } \hbox{
    \includegraphics[scale=0.5]{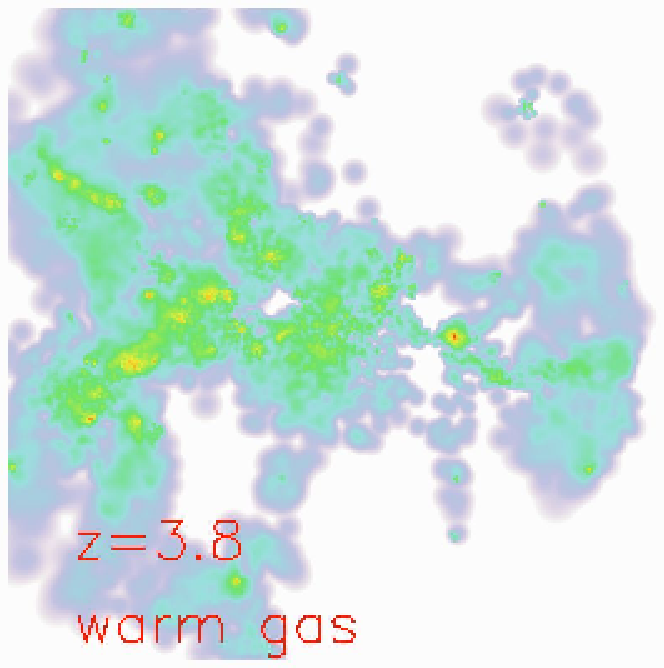}
    \includegraphics[scale=0.5]{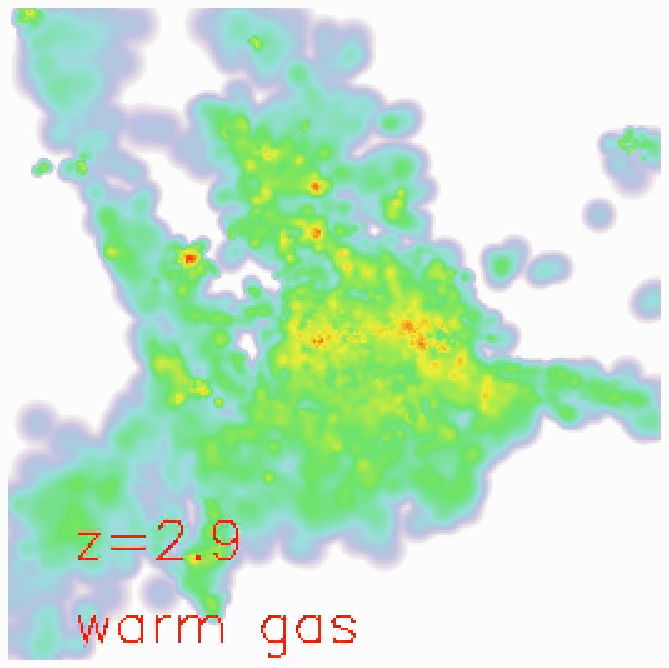}
    \includegraphics[scale=0.5]{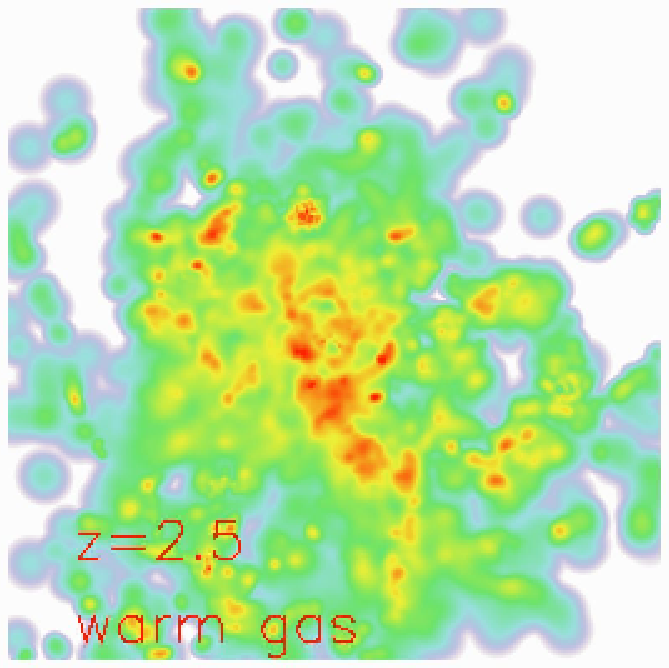} } \hbox{
    \includegraphics[scale=0.5]{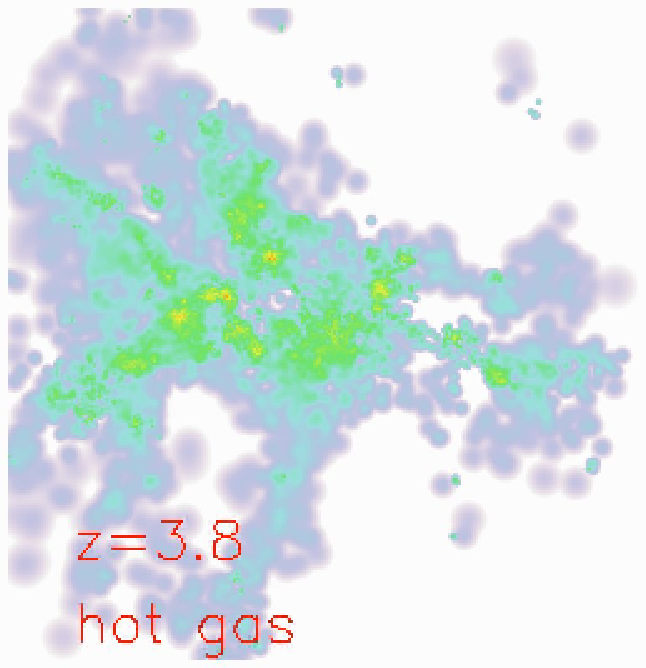}
    \includegraphics[scale=0.5]{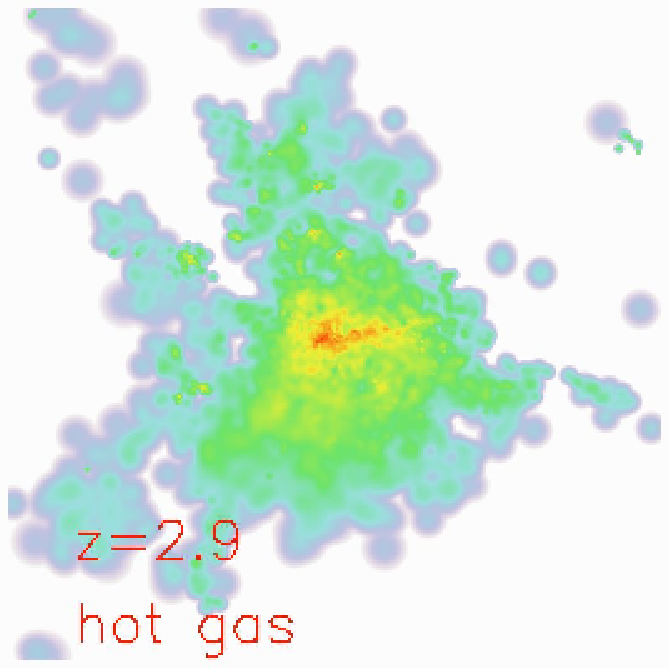}
    \includegraphics[scale=0.5]{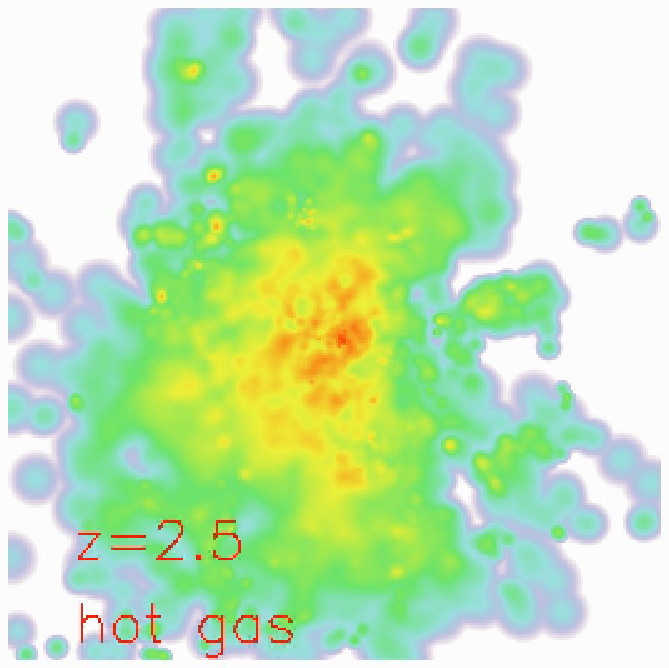} } }
\caption{Projected density maps of gas accreting on the {\it galaxy} between
  $z=2.5$ and $z=2$, for the R1 simulation. Different rows correspond to the
  three channels considered: cold (upper row), warm (middle row), and hot
  (lower row). Different columns correspond to different redshift: $z=3.8$
  (left column), $z=2.9$ (middle column), and $z=2.5$ (right column). The scale
  of the maps are 960, 480, and 240 h$^{-1}$Mpc comoving for the three
  columns. The color scale is logarithmic from 1.0 to $10^6$ M$_\odot$ h
  pc$^2$ comoving (from pale pink to red).
\label{fig:faces}}
\end{figure*}

Figure~\ref{fig:faces} shows density maps of gas which accreted onto
the {\it galaxy} between redshift $z=2.5$ and $z=2.0$. We
show the gas at previous redshifts: $z=3.8$ (left column), $z=2.9$
(middle column), and $z=2.5$ (right column), for the R1
simulation. Maps are centered on the most massive progenitor at
the corresponding redshift.  The upper, middle and lower rows are for
the {\it cold}, {\it warm} and {\it hot} channels, respectively. The
Figure shows that the cold gas is characterized by a very clumpy
distribution, while the hot gas is always diffused around the
most massive progenitor.  Interestingly, warm gas is more diffuse than
cold gas, but it surrounds the cold clumps.  This suggests that
the infalling cold clumps are strongly star forming, and that they
heat and eject part of their gas while infalling towards the central
galaxy. 

This is confirmed by the analysis of the multi-phase nature of the accreting
gas: SN thermal feedback is effective at heating multi-phase particles. So, by
computing the fraction of gas particles that were multi-phase when they
reached their maximum temperature, one can disentangle the influence of SN
heating from gravitational shock heating. For the R1 simulation, averaging
over the entire redshift range (from $z=8.3$ to $z=0$), we find that when
accreting gas reaches its maximum temperature, $\sim 97$\% of it is
multi-phase and star forming in the {\it cold} channel. The corresponding
fractions for the {\it warm} and {\it hot} channels are $\sim 66$\% and $\sim
8$\%, respectively. This suggests that gas in the {\it warm} channel is heated
by SN feedback, while gas in the {\it hot} channel is heated by gravitational
shocks.

To better illustrate the role of effective thermal feedback
we run simulation R1 using the widely used
effective model of
Springel \& Hernquist (SH, 2003) for star formation, with kinetic feedback and
velocity $v_w=340$ km s$^{-1}$
\footnote{We decouple wind particles from ambient gas for
20 kpc or until density drops to one half of SF threshold}. 
The resulting accretion rates are shown in the lower panels of Figure
\ref{fig:accr_channels}: 
the contribution of the {\it warm}
channel to accretion on the galaxy becomes negligible, while gas accretion in
the {\it cold} channel increases.
This confirms that simulations with uneffective thermal feedback
  are unable to heat cold flows into the warm channel, even when they
  produce significant outflows.

Finally, for each channel, we analyse the {\it clumpiness} of gas accretion, 
both on the {\it halo} and on the {\it galaxy} by determining if its particles
belonged to an accreting satellite of were accreted smoothly. 

\begin{figure*}
\hspace{0.5truecm}
\includegraphics[scale=0.85,angle=-90]{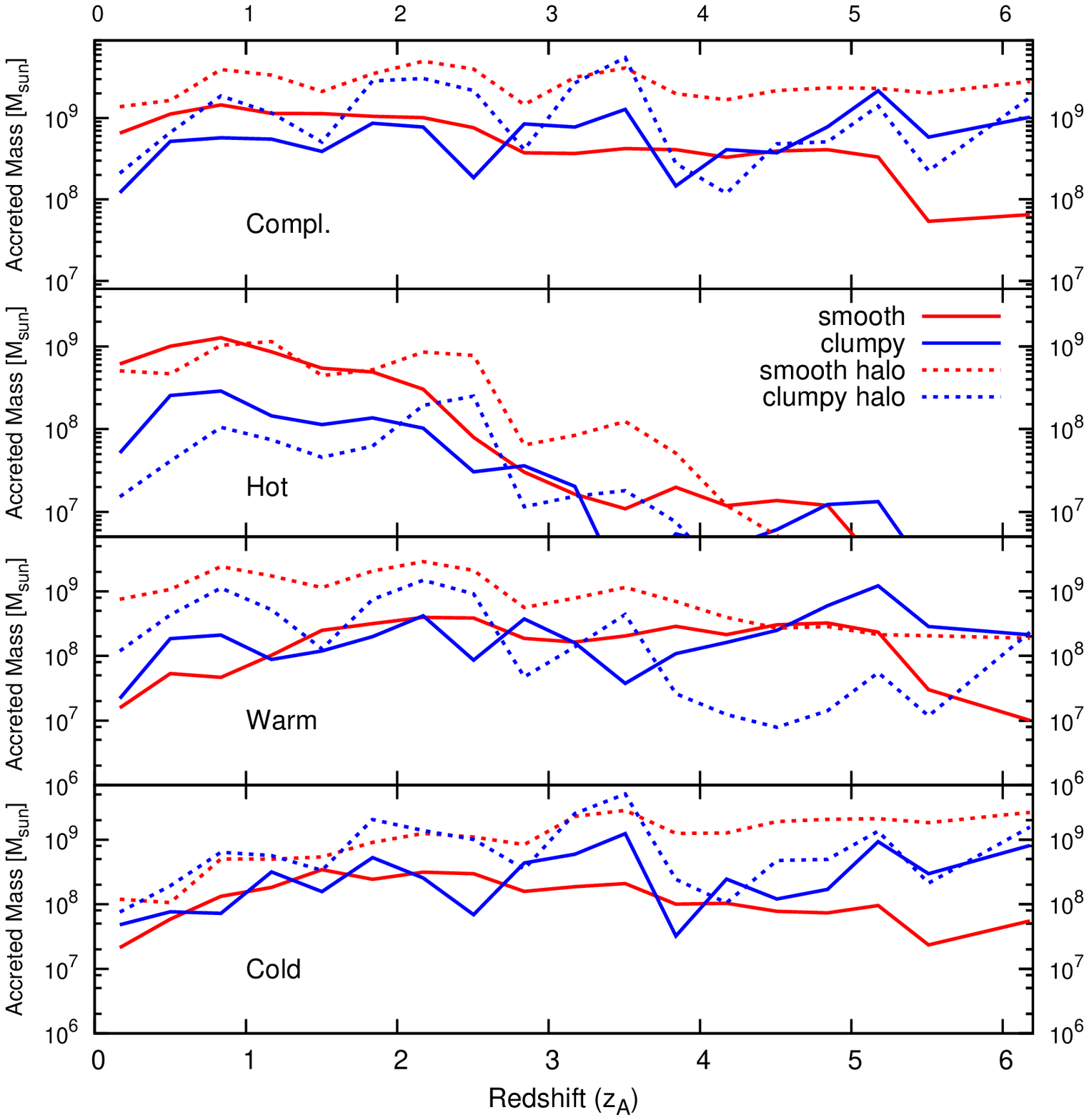}
\caption{Smooth vs clumpy accretion. Solid lines refer to accretion on
  the {\it galaxy}, while dashed lines are for accretion on the {\it
    halo}. Blue and red lines correspond to {\it clumpy} and {\it
    smooth} accretion, respectively. The upper panel shows the total
  accretion, while the three channels considered are shown in
  the lower panels: from the 2$^{nd}$ upper to lower panel, accretions
  for {\it hot}, {\it warm} and {\it cold} channels. Similar results
  are found for lower resolutions (see text).
\label{fig:smoothclumpy}
}
\end{figure*}

Results are shown in Figure~\ref{fig:smoothclumpy}.  For accretion on the halo,
we find that {\it total} gas is mainly smooth (as in, e.g, Brooks et
al. 2003, van de Voort et al. 2011). At $z=0$, $69$ per
cent of the gas has been accreted smoothly in the R1 simulation.
Smooth accretion dominates the {\it hot} channel,
while in the {\it cold} one, smooth and clumpy accretion contribute
equally to the total accretion rate: at $z=0$ smooth accretion
accounts for only 49 per cent of the overall cold accretion.

Also the total accretion rate on the galaxy is mainly smooth, $\approx
67$ per cent for R1, and hot accretion shows the same trend found for
the accretion on the halos. We find a significant difference, however,
in the {\it cold} channel.  In fact, this channel is dominated by
clumpy accretion, with 60 per cent of the gas accreted in this
way. The gas accreted through the {\it warm} channel is also
prevalently clumpy.

We stress that the clumps that contribute to the cold and warm
channels have a significant ($>$ 85 per cent) DM fraction, ruling out
possible numerical problems of the SPH technique that can, in certain
conditions, produce ``cold-gas only'' spurious blobs.

\section{Discussion and Conclusions}
\label{sec:concl}

We analyzed the gas accretion rate in cosmological numerical
simulations of a galaxy-size halo. We take advantage of our new MUPPI
star formation and feedback model that provides effective {\it
  thermal} feedback 
and use different resolution levels to test the robustness of our results.

We  distinguished between  gas accretion  onto the  halo (defined  as  all the
matter included  in the redshift-dependant virial radius),  and accretion onto
the central galaxy  (defined as all the material included  within one tenth of
the virial radius). We traced  back all gas particles in our simulations,
and defined a gas particle to be accreted at redshift $z_n$ if it was not part
of  the main  progenitor  (or of  its central  galaxy)  of our  $z=0$ halo  at
$z_{n-1} >  z_n$, but became part of  it at $z_n$. We  defined three accretion
channels: a cold  one, including gas particles whose  maximum past temperature
was lower than $2.5 \times 10^5$ K, a hot channel, that includes gas particles
with maximum  past temperature larger  than $10^6$K, and an  intermediate {\it
  warm} channel.

Our results can be summarized as follows:

\begin{itemize}
\item The {\it warm} accretion channel becomes as important as the cold one at
  high redshift, when considering the accretion on the central {\it
    galaxy}. This trend is robust against numerical resolution.
\item Gas particles in the {\it warm} channel are heated by SN
  feedback. This is confirmed by the analysis of the multi-phase
  nature of the gas, and by the fact that a
  simulation of the same halo carried out with a model that does not
  include effective thermal feedback shows negligible accretion rates
  through this channel.
\end{itemize}

Regarding the accretion on the halo, our findings confirm previous
studies: the `cold' accretion mode on the halo is important at high
redshift, while the `hot' accretion mode becomes dominant at lower
redshift. However, accretion rates onto the halo are different from
accretion rates onto the central galaxy.  We find that a significant
amount of gas is accreted on the galaxy through a {\it warm}
channel. The gas is in cold clumps when entering the halo, and becomes
warm because of the SN feedback originating from star formation within
the clumps.

Van de Voort et al. (2011) too analyzed the clumpiness of the accreted gas,
but they included all clumps with mass ratio smaller than 1:10 in
their smooth component. These clumps are, however, very dense and star-forming,
and effectively heat the cold gas to the warm channel.

Oppenheimer et al. (2010) found that a large fraction of low-z star
formation experienced a wind phase in the past.  However, they studied
the effect of {\it kinetic} feedback, and showed that this channel
becomes important below $z=2$.  Their results are thus complementary
to those shown in this letter.

The picture emerging from our analysis can be sketched as follow: at high
redshift, accreting gas is mainly cold, partly clumpy and partly
smooth. Not all of the smooth cold gas reaches the central part of the
halos because, behaving as non-collisional material, cold, un-shocked gas is
not able to loose orbital energy and angular momentum. Thus, only gas streams
with very low impact 
parameters can funnel gas to the central galaxy, contributing to its
star formation. Cold clumps suffer dynamical friction and spiral
towards the center. Since they are very dense, their star formation
rate is high, and part of the gas they contain is heated by SN
feedback, and is accreted onto the galaxy as {\it warm}, slightly less
clumpy, material.  We do not ascribe these findings to specific
features of our sub-resolution model, and argue that {\it any}
effective thermal feedback scheme would give similar results.
In this picture, the time needed by the cold gas to
reach the central galaxy is neither the free fall time nor the cooling
time. 
Smooth cold accreting gas can only reach the galaxy if the impact parameter is
small. In contrast, cold clumps are subject to dynamical friction so they can
reach the central galaxy even with non negligible impact parameters, over a
time-scale that is determined by dynamical friction.

However, the exact quantification of warm flows may well depend
  on model details. In particular, the entrainment of cold gas into
  the multi-phase particle may lead to some overestimation of the
  amount of warm flows. We will address this
  issue in future work.

It is widely accepted that stellar feedback plays an important role in
determining the physical properties of galaxies. It remains, however, unclear
if the nature of feedback is predominantly kinetic or thermal. Our results
demonstrate that stellar feedback strongly affects the termodynamical state of
the gas accreting onto galaxies. In particular, we argue that a mode of feedback
that is primarily thermal would produce significant amount of warm gas 
accreting onto galaxies at high redshift. 
This has noticeable consequences on the observability of gas
  flows infalling onto high redshift galaxies, and 
  may explain, at least in part, why cold flows have been difficult to
  detect, to date.

\acknowledgments 

We acknowledge useful discussions with Matteo Viel. Our simulations were
carried out at the ``Centro Interuniversitario del Nord-Est per il Calcolo
Elettronico'' (CINECA, Bologna), with CPU time assigned under two INAF/CINECA
grants. The R2 simulation was performed thanks to an INAF-CINECA numerical
Key-Project. GDL acknowledges financial support from the European Research
Council under the European Community's Seventh Framework Programme
(FP7/2007-2013)/ERC grant agreement n. 202781. This work has been partially
supported by the European Commission FP7 Marie Curie Initial Training Network
CosmoComp (PITN-GA-2009-238356), by PRIN-INAF-2009 Grant "Towards an Italian
Network for Computational Cosmology", by PRIN-MIUR09 ``Tracing the growth of
structures in the Universe'', and by the INFN PD51 grant.


\begin{thebibliography}{}
\bibitem[Blitz \& Rosolovski (2006)]{BlitzRosolowski06}Blitz, L.,  Rosolowsky, E. 2006, \apj, 650, 933
\bibitem[Benson \& Bower (2011)]{BensonBower11}Benson, A.J, Bower, R. 2011,  \mnras, 410, 2653
\bibitem[Bertschinger (1985)]{Bertshinger85}Bertchinger, E. 1985, \apjs, 58, 39
\bibitem[Binney (1977)]{Binney77}Binney, J. 1977, \apj, 215, 483
\bibitem[Binrboim (2003)]{Binrboim03}Birnboim, Y., Dekel, A. 2003, \mnras, 345, 349
\bibitem[Brooks et al. (2009)]{Brooks09}Brooks, A.M., Governato, F., Quinn, T. Brook, C.,  Wadsley, J. 2009, \apj, 694, 396
\bibitem[Evrard (1990)]{Evrard90}Evrard, A.E. 1990, \apj, 363, 349
\bibitem[Forcada-Miro \& White (1977)]{Forcada77}Forcada-Miro, M.I.,  White,  S.D.M. 1997, arXiv:astro-ph/9712204
\bibitem[Haardt \& Madau (1996)]{Haardt96}Haardt, F., Madau, P. 1996, \apj,  461, 20
\bibitem[Keres et al. (2005)]{Keres05}Keres, D., Katz, N., Weinberg, D.H., Dav\'e, R.  2005, \mnras, 363, 2
\bibitem[Keres et al. (2009)]{Keres09}Keres, D., Katz, N., Fardal, M., Dav\'e, R.,  Weinberg, D.H 2009, \mnras, 395, 160
\bibitem[Murante et al (2010)]{M10} Murante, G., Monaco, P, Giovalli,  M, Borgani, S., Diaferio, A. 2010, \mnras, 405, 1491
\bibitem[Okvirk et al. (2008)]{Okvirk08}Ocvirk, P., Pichon, C., Teyssier, R  2008, \mnras,  390, 1326
\bibitem[Oppenheimer et al. (2010)]{Oppenheimer10} Oppenheimer, B.D.,
  Dav\'e, R., Keres, D., Fardal, M., Katz, N., Kollmeier, J.A.,
  Weinber,  D.H. 2010, \mnras, 406, 2325
\bibitem[Rees \& Ostriker (1977)]{ReesOstriker77}Rees, M.J., Ostriker, J.P.  1977, \mnras, 179, 541
\bibitem[Scannapieco et al. (2009)]{Scannapieco09}Scannapieco, C.,
  White, S.D.M.,  Springel, V., Tissera, P.B. 2009, \mnras, 396, 696
\bibitem[Scannapieco et al. (2011)]{Scannapieco11}Scannapieco, C., et
  al 2011, \mnras, submitted, arXiv:1112.0317
\bibitem[Springel et al. (2001)]{Springel01}Springel, V., White, S.D.M, Tormen,  G., Kauffmann, G. 2001, \mnras, 328, 726
\bibitem[Springel \& Hernquist (2003)]{SH03}Springel, V. Hernquist, L  2003, \mnras, 339, 289
\bibitem[Schaye et al. (2010)]{Schaye10}Schaye, J., Dalla Vecchia, C., Booth,  C.M., Wiersma, R.P.C., Theuns, T., Haas, M.R., Bertone, S., Duffy, A.R.,  McCarthy, I.G., van de Voort, F. 2010, \mnras, 402, 1536
\bibitem[Springel (2005)]{Springel05}Springel, V. 2005, \mnras, 364, 1105
\bibitem[Stoehr (2002)]{Felix02}Stoehr, F. White, S.D.M., Tormen, G.,
  Springel,  V. 2002, \mnras, 335, 84
\bibitem[van de Voort et al. (2011)]{vandevoort11}van de Voort, F.,
  Schaye, J., Booth, C. M., Haas, M.R., Dalla Vecchia, C. 2011,\mnras,414,2458
\bibitem[White \& Rees (1978)]{WH78}White, S.D.M., Rees, M.J. 1978, \mnras,  183, 341
\end{thebibliography}
\bibliographystyle{plainnat}

\clearpage

\end{document}